\documentstyle[12pt]{article}
\newcounter{mycount}

\newcommand{\be}{\begin{eqnarray}}
\newcommand{\ee}{\end{eqnarray}}

\newcommand\ie {{\it i.e. }}
\newcommand\eg {{\it e.g. }}

\newcommand\vx{\vec{x}}
\newcommand\ketk{|k;\vx_i\rangle}
\newcommand\ketkp{|k+1;\vx_i\rangle}
\newcommand\ketl{|l;\vx_i\rangle}
\newcommand\ketlm{|l-1;\vx_i\rangle}
\newcommand\etal{{\it et. al.}}

\newcommand\noi{\noindent}

\begin{document}

\bibliographystyle{nphys}

\centerline{\Large\bf ANYONS ON HIGHER GENUS SURFACES - }
\vskip 2mm
\centerline{\Large\bf a constructive approach}
\vspace* {-30 mm}
\begin{flushright}  USITP-93-11 \\
May 1993
\end{flushright}
\vskip 30mm
\centerline{T. H. Hansson$^{\dagger}$, Anders Karlhede$^{\dagger}$
and Erik Westerberg}
\vskip 4mm
\centerline {hansson@vana.physto.se  ak@vana.physto.se wes@vana.physto.se }
\centerline{Department of Physics  }
\centerline{University of Stockholm } \centerline{ Box 6730 } \centerline {
S-113 85 Stockholm} \centerline{Sweden }
\vskip 2.0cm
\newcommand \sss {\mbox{ $<\overline{s}s>$} }
\def\fk{\mbox{ $f_K$} }
\centerline{\bf ABSTRACT}
\vskip 3mm
We reconsider the problem of anyons on higher genus surfaces by
embedding them in three dimensional space. From a concrete
realization based on three dimensional flux tubes bound to charges moving
on the
surface, we explicitly derive all the representations of
the spinning braid group. The component structure of the wave functions arises
from winding the flux tubes around the handles. We also argue
that the anyons in our construction
must fulfil the generalized spin-statistics relation.

\vfil
\vskip 5mm
\noindent
$^{\dagger}$Supported by the Swedish Natural Science Research Council
\eject

Anyons, \ie particles with fractional statistics can exist in two space
dimensions\cite{lein1}. The statistical angle, $\theta$, which by
definition is the
phase acquired by the wave function under the clock-wise interchange of two
particles,
is an arbitrary real number, defined mod $2\pi$, if the particles move on an
infinite
plane. When the particles move on surfaces with more complicated topology,
$\theta
/\pi$ is restricted to special rational values. On the sphere, an
infinitesimal loop traced by one of the
particles can also be thought of as a big loop enclosing all the
other particles. This implies a relation between $\theta$, the number of
particles and the spin of the particles. On
higher genus surfaces the situation is more complicated and was clarified
only when Einarsson realized that the relevant representations of the braid
group on the torus are multi-dimensional, and the wave functions hence have
several components \cite{eina1}.
 A fair amount of work concerning anyons on the
torus exists, in particular it has been shown how to recover the
component structure of the wave functions
in the Chern-Simons (CS) approach \cite{ieng1,lech1,fayy1}.  Imbo and
March-Russel\cite{imbo1} and Einarsson\cite{eina2} generalized the
results for the torus
to surfaces of arbritrary genus. The relations between the
number of particles, $N$, the
dimensionality of the representation, $d$, (\ie the number of
components of the  wave
function) and the statistics, $\theta/\pi = p/q$, depend on the spin of
the anyons
\cite{wilc1,bala1,bala2}. This is because the coupling of the spin to the
curvature of the
surface gives a phase change in the wave function when the particle moves.
Einarsson pointed out that to get agreement with the results from Chern-Simons
theory, one must assume that the anyons carry a spin  given by the generalized
spin-statistics relation $s=\theta/2\pi$. This is  consistent with the explicit
calculations of spin in CS theories\cite{hage1} and Maxwell-CS theories
\cite{hans2,artz1}. The relations pertinent to  anyons with
spin $s=\theta/2\pi$
on an orientable genus $g$ surface, derived in refs. \cite{imbo1} and
\cite{eina2}, read \be  \theta /\pi = M/N = p/q \ \ ,
\ \ \ \ \ d=q^g \ \ \ \ \ \
\ , 
  \label{eq:srel}
\ee
where $M$ is an arbitrary integer and $p$ and $q$ are relative prime
(defined by $p/q=M/N$).
Thus for a given number of anyons on a given surface, the only freedom is
the integer $M$,
which uniquely determines the statistics and the number of components of the
wave function.

The possible values for the statistical angle $\theta$ in (1) can be obtained
as follows, see
\eg \cite{eina3}. Embed the surface with $N$ bosons in three dimensions,
and attach a three
dimensional tube of  magnetic flux $2\theta$ to each
particle\footnote{We put $c=\hbar =e=1$ so that a unit
quantum of flux is $2\pi$.}. One such  tube crosses the
surface at the position  of each  particle, and there is no magnetic flux
flowing through the
surface anywhere else, see Fig. 1. By the Aharonov-Bohm effect, this flux
turns the particles into
anyons of statistics $\theta$.  By Dirac's quantization condition, they must
originate in an integer
number, $M$ say, of magnetic monopoles inside the surface. The flux
through each particle is therefore
$2\theta=2\pi M/N=2\pi p/q$.

In this letter, we derive all representation of spinning anyons from this
 three dimensional realization.
The component structure of the wave functions, as well as the explicit
representations for the generators of the braid group,  are derived from
simple physical
properties of the three dimensional realization. For simplicity we imagine
that the flux tubes end in $M$
anti-monopoles outside the torus and that all 2$M$ monopoles are kept fixed
in space. The dynamics
of the fluxes and monopoles will not be relevant for our arguments. The
details of how the flux tubes are attached to
the particles are not important, but below we will comment on how this can
be achieved via
a constraint in the Hamiltonian.

The principle behind our construction is that a quantum state of the anyons is
specified  not only by the particle positions, but also by  the values of
non-contractible
Wilson loops, $e^{i\oint A}$, on the surface. We  give the details of the
construction for the torus;  the generalization
to higher genus surfaces will be obvious. Let $\rho_i$ be an operator that
moves particle {\it i} around the loop
$\rho_i$, see Fig. 1.  We assume that the eigenvalues of the quantum
mechanical operator $\rho_i$ are the possible
values the Wilson loop measuring the (three dimensional) flux through
$\rho_i$. The operator $\tau_j$ is defined
analogously. Classically, when
 an anyon moves on the torus, a three dimensional flux  tube is dragged
along.  Moving the anyon once around
$\rho_i$ restores the particle configuration but wraps a tube of $p/q$
flux quanta  around inside the torus, thus
changing the flux through $\tau_j$. Since we have assumed that the
eigenvalues of the operators are the Wilson
loops, it follows that the quantum mechanical operators $\rho_i$ and
$\tau_j$ do not commute. Consider an eigenstate,
$\ketk_\tau$,  of the operator $\tau_1$. Acting on this state with
$\rho_1$, gives a new state, $\ketkp_\tau$, since
the flux through $\tau_1$ has changed. Repeated actions of $\rho_1$
gives new states until, after $q$ steps, an
integer number of flux quanta, $p$,  have been added.  The initial state
has then been retrieved, possibly
up to a phase:
 $\rho_1^q\ketk_\tau =e^{iq\eta}\ketk_\tau$. Choosing the obvious
phase convention we have,
\be
\rho_1 \ketk_\tau  =
e^{i\eta}\ketkp_\tau \ \ , k=0,2,...q-1 & {\rm and} &
|q;\vx_i\rangle_\tau =|0;\vx_i\rangle_\tau \ \ .
 \ee
The phase $e^{i\eta}$ corresponds to  a constant  flux flowing
through the hole of the torus. The $q$ states are the $q$ components of
the wave function on the torus. Taking the
$j^{th}$ particle around the loop $\rho_j$ will have the same effect as
$\rho_1$ except that the
enclosed flux will also get a contribution from the enclosed $(j-1)$ flux
tubes connected to
the particles. Keeping track of signs we get
$\rho_j = e^{-2i\theta (j-1)}\rho_1$.
The eigenstates to the $\rho$'s are the superpositions
\be
\ketl_\rho & = & \frac{1}{\sqrt{q}}\sum_k e^{-2ikl\theta}\ketk_\tau
\ \ \ \ , \ \ \ \ \  \
l=0,1,...q-1, \ee
with eigenvalues $e^{i\eta}e^{2il\theta}$ measuring the enclosed flux.

Next consider the action of the generator $\tau_1$. In
analogy  to the $\rho$'s, the $\tau$'s  decrease the flux as measured
by the $\rho$'s by
$2\theta$:
\be
\tau_1 \ketl_\rho =  e^{i\gamma}\ketlm_\rho \ \ \ ,
\ee
where the phase $e^{i\gamma}$ stems from a constant flux flowing inside
the torus.
{}From Fig. 1 it is also clear that
$\tau_j  =  e^{2i\theta (j-1)}\tau_1$ since the enclosed flux decreases by
$ 2\theta(j-1)$ due to
the encircled $(j-1)$ flux tubes. The eigenstates of $\tau_j$ are the states
$\ketk_\tau$, with eigenvalues $e^{i\gamma}e^{2ik\theta}$. As promised, the
eigenvalue measures the flux through
$\tau_1$ since $k2\theta$ is the flux added when acting with $(\rho_j)^k$, and
$\gamma$ is the "external" flux.  (The picture in Fig.1 is classical; it does
not
correspond to a $\tau$ (or $\rho$) eigenstate.)

Finally, we note that the local interchange $\sigma_i$ of two
particles does not change the component structure but only causes the wave
function to change with
the familiar statistical phase $e^{i\theta}$ so that
\be
\sigma_i \ketk_\tau \ = \ e^{i\theta}\ketk_\tau & {\rm and} &
\sigma_i \ketl_\rho \ = \ e^{i\theta}\ketl_\rho \ \ .
\ee
The representation of $\tau_j$, $\rho_j$ and $\sigma_j$ in either the
$\{\ketk_\rho\}$ or the $\{\ketl_\tau\}$ basis reproduces exactly the braid
group representations found by Einarsson for anyons on the torus \cite{eina1}.

We can now proceed to construct  a path
integral representation for the propagator, or partition function, for the
anyon
system using the  conventional recipe: 1. Divide the paths according to
homotopy
classes; 2. Perform the sum over  paths in each class separately; 3.  Multiply
each contribution with an appropriate phase-factor  and sum over all
classes. Note
that for a path to connect the states $|k;\vec x_i\rangle$ and  $|k';\vec
x_i'\rangle$,  the particles must wind altogether $k-k'$ modulo $q$
turns in the
$\rho$-direction.  Also, when calculating the
relative phases appropriate to the different homotopy classes, one must
consistently use one representation for the braid group generators $\tau_j$,
$\rho_j$ and $\sigma_j$.

The generalization to higher genus surfaces is trivially obtained by
representing
the genus $g$ surface as a sphere with $g$ handles. The statistics is
independent
of genus since  $M$ monopoles inside give $M/N=p/q$ flux quanta per
particle and
hence statistics $\theta/\pi=p/q$.   Particle transport around different
handles
commute (since winding around handle A does not affect the Wilson loops  around
handle B) while the generators on the same handle behave as on the torus. There
are hence $q$ inequivalent windings on each handle and all in all $q^g$
components
of the wave function and the relations (1) are recovered.

Three comments are in order. First, the number of components, for the
torus, is $q$ and not $q^2$
which might naively have been guessed since we can wind flux tubes around
both holes of  the
torus. This is because the component structure is not caused by the total
flux through the hole but by the
flux added when acting with $\rho_i$ or $\tau_i$ (\ie when moving one
particle around). Since
$\rho$ and $\tau$ do not commute, only one of them will generate components.
If we diagonalize
$\rho$, then $\tau$ generates $q$ components and vice versa.
Second, although the three dimensional embedding distinguishes the two holes
in the torus, the
two dimensional physics is symmetric; which hole one adds flux through just
corresponds to a
choice of basis.
Third, we reproduced the relations corresponding to anyons with fractional
spin in accordance with
the generalized spin-statistics relation. Although our understanding of this
is not complete, we
shall argue below that this is no coincidence.

We   stressed that the details of how flux is bound to the charges is of no
relevance for
the argument about the component structure. From that point of view we
could stop here. It is, however,
interesting to ask whether the flux tube picture can be made dynamical,
and without going into any
details we will argue two points: 1. Any consistent dynamical scheme will
imply phases corresponding to particle
transport in accordance with eqs. (2), (4) and (5). 2. Any flux tube-charge
composite will carry
fractional spin in accordance with the generalized spin-statistics relation.

One way to bind flux to charge on the surface is to use a  constraint as in
ref.
\cite{zhan1}. On a plane it is well understood how this construction works:
Describe the $N$ charged particles in the xy-plane by a path-integral and
couple
them minimally to the gaugepotential $\vec a = (A_x, A_y)$ which is the
restriction of the 3-d vectorpotential $\vec A$, which describes the flux
tubes,
to the xy-plane.  Next, tie the charges to the flux tubes by introducing a
delta-functional via the identity $1=\int {\cal D}b \, \delta[( b(\vec x) -
2\theta\sum_{i=1}^N\delta^2(\vec x_i)]$ in the path integral  description
of the
system. This constraint is then exponentiated using a Lagrange multiplier field
$a_0$, and as shown in \cite{zhan1}, the resulting $a_0 b$ term
(where $b=B_z$) in
the action is nothing but a Chern-Simons (CS) term in Coulomb gauge.

On the torus, everything works locally as on the plane. After
exponentiating the
constraint one has a path integral over $b$ and $a_0$ and $a_0 b$ in
the action.
However, this is not equivalent to the full CS action on the torus; the
non-trivial Wilson loops are missing since we integrate over $b$ only. When
quantizing the full CS action, \ie integrating over all gauge fields
$a_\mu$, then
the two non-trivial Wilson loops, corresponding to $\rho$ and $\tau$, are
conjugate variables and give a single quantum degree of freedom
\cite{poly7,fayy1}. The  component structure  explained above exactly
corresponds
to a single quantum degree of freedom with the same spectrum as the Wilson loop
mode, and we believe that the two descriptions are indeed equivalent. Our
construction then, is a ''mixed'' one: The statistical phase is implemented by
flux tubes (or gauge fields), and the wave functions are single valued but have
several components associated with $\rho$ and $\tau$. In the description
of anyons
with a full CS term \cite{fayy1}, the Wilson loop mode replaces the component
structure.

As discussed in  \cite{hans2} and \cite{gold3}, the composites are anyons with
statistical angle $\theta$. Note that this is not self evident. Naively
one would
expect a phase $2\theta$   since there are equally big contributions from the
charge moving relative to the flux, and the flux relative to the
charge.\footnote{Such composites were named ''cyons'' by
Goldhaber\cite{gold1}. }
However, there is a contribution to the phase from the CS term, \ie from the
constraint, that cancels half of the naive contribution.  Since $a_i$ is the
restriction of $A_i$ to the surface, this proves the assertion that  the
wave-functions pick up  phases equal to the Aharonov-Bohm phase
corresponding to a
unit test charge following the same trajectory. In particular this
means that the
statistical phase corresponding to interchanging two particles will
be $\theta$.

The simplest way to see that anyons in the CS theory carry spin is to calculate
the corresponding spinfactors, as explained in refs. \cite{poly1}, \cite{grun1}
and \cite{grun3}. Another way is to directly evaluate the canonical angular
momentum from the Lagrangian. Both methods give results in accordance with the
spin-statistics connection.\footnote{  The calculations in \cite{artz1}
show that
it is not crucial that the anyons are point like, and it is demonstrated
that the
spin-statistics connection holds for a large class of Lagrangians.}
A third way to understand the spin is by appealing to the topological
proof of the spin-statistics theorem given by  Balachandran
\etal\cite{bala1,bala2}. This proof does not require Lorentz invariance, but is
based on the following assumptions: {\it i})  A continuous, spacelike,
''frame''
can be associated with each point on the worldline of a  particle. {\it ii})
There
exist antiparticles that carry mirror-reflected frames. {\it iii})
Particle-antiparticle pairs can be created and annihilated only if their frames
are aligned (see Fig. 2b). In our flux tube picture there is  a very natural
way
to attach frames to the anyons. Define $\hat{e}_1$ to be the charge of
the particle times the projection of  the unit tangent vector to the
magnetic flux tube at the position of the anyon ($\hat{e}_1$ normalized to unit
length) and $\hat{e}_2$ to be the (normalized) cross-product between the
tangent vector and $\hat{e}_1$ (Fig. 2a). It is clear that the frame of an
anti-anyon will be the mirror image of the frame of an anyon and also that the
frames are aligned as required in any annihilation/creation process (Fig. 2b).
There is a difficulty in that the frame is not well defined if the flux tube is
perpendicular to the surface, but if we modify the construction and exclude
this
possibility then the flux tube picture gives a very nice concrete
realization of
the assumptions going into the topological proof of the spin-statistics
theorem.

Again we must generalize the treatment to a curved space. This we have
not done in general, but we have
checked that in the case of a sphere the so called non-covariant, or
twist, form of the
spinfactor\cite{poly1,grun3} picks up precisely that dependence on the
curvature that is necessary for consistency
with the braid group analysis of Einarsson. Also, it seems to us that the
argument based on the topological proof of the spin-statistics theorem
generalizes directly to a curved surface.
\vskip 3mm\noi
{\bf Acknowledgement:} We thank Ansar Fayyazuddin for many helpful discussions
that forced us to sharpen our arguments. We also thank Martin Ro\v cek for
discussions.


\begin{thebibliography}{10}

\bibitem{lein1}
{J. M. Leinaas,} and J. Myrheim,
\newblock {\it Nuovo Cimento}, {\bf 37B} (1977)1.

\bibitem{eina1}
T. Einarsson,
\newblock {\it Phys. Rev. Lett.}, {\bf 64} (1990)1995.

\bibitem{ieng1}
R. Iengo and K. Lechner,
\newblock {\it Nucl. Phys.}, {\bf B346} (1990)551.

\bibitem{lech1}
K. Lechner,
\newblock {\it Anyon Physics on the Torus},
\newblock Thesis, 1991.

\bibitem{fayy1}
A. Fayyazuddin,
\newblock {\it On the origin of multi-component anyon wave functions},
\newblock Preprint, Stockholm University, USITP-92-15, {\em Nucl. Phys.} {\bf
  B}, in press, 1992.

\bibitem{imbo1}
T.~D. Imbo and J. March-Russel,
\newblock {\it Phys. Lett.}, {\bf B 252} (1990)84.

\bibitem{eina2}
T. Einarsson,
\newblock {\it Mod. Phys. Lett.}, {\bf B5} (1991)675.

\bibitem{wilc1}
F. Wilczek,
\newblock {\it Phys. Rev. Lett.}, {\bf 49} (1982)957.

\bibitem{bala1}
A.P. Balachandran, A. Daughton, Z.-C. Gu, G. Marmo, R.~D. Sorkin, and A.~M.
  Srivastava,
\newblock {\it Mod. Phys. Lett.}, {\bf A5} (1990)1575.

\bibitem{bala2}
A.P. Balachandran, A. Daughton, Z.-C. Gu, G. Marmo, R.~D. Sorkin, and A.~M.
  Srivastava,
\newblock {\it Spin-Statistics theorems with no relativity or field theory},
\newblock Preprint, Syracuse University Preprint SU-4228-433, May 1990.

\bibitem{hage1}
C.~R. Hagen,
\newblock {\it Phys. Rev.}, {\bf D31} (1985)2135.

\bibitem{hans2}
{T. H. Hansson,}~{M. Ro\v cek,} {I. Zahed,} and S.~C. Zhang,
\newblock {\it Phys. Lett.}, {\bf B214} (1988)475.

\bibitem{artz1}
A.~Karlhede S.~Artz, T.~H.~Hansson and T. Staab,
\newblock {\it Phys. Lett.}, {\bf B 267} (1991)389.

\bibitem{eina3}
T. Einarsson,
\newblock {\it Anyons and Antiferromagnets:Two Two-dimensional Topics},
\newblock Thesis, Chalmers University of Technology, G\" oteborg 1992.

\bibitem{zhan1}
{S-C. Zhang,} {T. H. Hansson} and S. Kivelson,
\newblock {\it Phys. Rev. Lett.}, {\bf 62} (1989)82.

\bibitem{poly7}
A.~P. Polychronakos,
\newblock {\it Ann. Phys.}, {\bf 203} (1990)231.

\bibitem{gold3}
A.~S. Goldhaber and R. Mackenzie,
\newblock {\it Phys. Lett.}, {\bf B214} (1988)471.

\bibitem{gold1}
A.~S. Goldhaber,
\newblock {\it Phys. Rev. Lett.}, {\bf 49} (1982)905.

\bibitem{poly1}
A.~M. Polyakov,
\newblock {\it Mod. Phys. Lett.}, {\bf 3} (1988)325.

\bibitem{grun1}
{J. Grundberg,}~{T.H. Hansson,} {A. Karlhede,} and U. Lindstr{\"o}m,
\newblock {\it Phys. Lett.}, {\bf B218} (1989)321.

\bibitem{grun3}
{J. Grundberg,} {T.H. Hansson,} and {A. Karlhede,},
\newblock {\it Nucl. Phys.}, {\bf B347} (1990)420.

\end{thebibliography}
\end{document}